\documentclass{PoS}

\usepackage{epsf}
\usepackage{amssymb}
\usepackage{amsmath}
\usepackage{amsfonts}
\usepackage{cite}
\usepackage[small]{caption2}
\usepackage[T1]{fontenc}
\usepackage{psfrag,epsfig,graphicx,graphics}
\usepackage{dcolumn}

\title{High-energy amplitudes and impact factors at next-to-leading order}

\ShortTitle{High-energy amplitudes and impact factors at next-to-leading order}

\author{\speaker{Giovanni Antonio Chirilli}%
         \\
         CPHT, \'Ecole Polytechnique, CNRS, 91128 Palaiseau Cedex, France \& LPT, Universit\'e Paris-Sud, CNRS, 91405 Orsay, France\\
        E-mail: \email{giovanni.chirilli@cpht.polytechnique.fr}}

\abstract{To study scattering amplitudes at high-energy, the T-product of two currents can
be expanded in terms of coefficient functions (impact factors) and matrix elements
of ``composite color dipoles'' made of Wilson line operators with rapidity cutoff preserving
conformal invariance. In the leading order, the high-energy evolution of color dipoles is governed by the
non-linear Balitsky-Kovchegov (BK) equation. To describe the high-energy amplitudes in the  next-to-leading order
(NLO) one needs to know the coefficient function (``impact factor'') and the evolution of
corresponding Wilson-line operators. Using the high-energy OPE, we find the next-to-leading order (NLO)
correction to the BK equation and calculate the impact factor for virtual photons in deep inelastic
scattering.}

\FullConference{XVIII International Workshop on Deep-Inelastic Scattering and
Related Subjects\\
                 April 19 - 23, 2010\\
                 Convitto della Calza, Firenze, Italy}

\newcommand{\dhd}{{\textstyle d}
\lower.03ex\hbox{\kern-0.40em$^{\scriptstyle-}$}\kern-0.08em{}}

\newcommand{\calu}{{\cal U}} 
 
\newcommand{\calz}{{\cal Z}}

\begin{document}

\section{Introduction}

Wilson line operators are the effective degrees of freedom for the description of high-energy scattering 
in gauge theories (for a review, see Ref. \cite{mobzor, Balitsky:2010jf}). Indeed, at high-energy (Regge limit) particles
move along their straight-line classical trajectory and the only quantum effect is the eikonal phase
factor acquired along this propagation path. In QCD, for fast quarks or gluons scattering off some
target, this eikonal phase factor is a Wilson line - an infinite gauge link ordered along the straight
line collinear to particle's velocity $n^\mu$:
\begin{equation}
U^\eta(x_\perp)={\rm Pexp}\Big\{ig\int_{-\infty}^\infty\!\!  du ~n_\mu 
~A^\mu(un+x_\perp)\Big\}~~~~
\label{defU}
\end{equation}
where $A_\mu$ is the gluon field of the target, $x_\perp$ is the transverse
position of the particle which remains unchanged throughout the collision, and the 
index $\eta$ is the rapidity of the particle. The high-energy behavior of QCD amplitudes can then be studied in the framework of the evolution
of color dipoles. We consider the small-x behavior of structure functions of deep inelastic
scattering (DIS): the virtual photon decomposes into quark and antiquark pair which
propagate along the straight lines separated by transverse distance and forms a color dipole -
two-Wilson-line operator:
\begin{equation}
\hat{\cal U}^\eta(x_\perp,y_\perp)=1-{1\over N_c}
{\rm tr}\{\hat{U}^\eta(x_\perp)\hat{U}^{\dagger\eta}(y_\perp)\}
\label{fla1}
\end{equation}
The energy dependence of the structure function is translated into the dependence of the color dipole on the rapidity $\eta$.
Although it appears to be more natural to restrict the rapidity by considering the Wilson line with the supporting line collinear 
to the velocity of the fast-moving particle, we choose to cut the rapidity integrals ``by hand'': 
the method of ``rigid cutoff'' in the longitudinal direction is technically simpler and more efficient in order to get the conformal results. 
Thus, the  small-x behavior of the structure functions is  governed by the 
rapidity evolution of color dipoles \cite{mu94,nnn}. 
At relatively high energies and for sufficiently small dipoles we can use the leading logarithmic approximation (LLA)
where  $ \alpha_s\ll 1,~ \alpha_s\ln x_B\sim 1$ and get the non-linear BK evolution equation for the color
dipoles \cite{npb96,yura}:
\begin{eqnarray}
&&\hspace{-26mm}
{d\over d\eta}~\hat{\cal U}^\eta(z_1,z_2)~=~
{\alpha_sN_c\over 2\pi^2}\!\int\!d^2z_3~ {z_{12}^2\over z_{13}^2z_{23}^2}
[\hat{\cal U}^\eta(z_1,z_3)+\hat{\cal U}^\eta(z_3,z_2))\nonumber\\
&&\hspace{30mm}-\hat{\cal U}^\eta(z_1,z_3)-\hat{\cal U}^\eta(z_1,z_3)\hat{\cal U}^\eta(z_3,z_2)]
\label{bk}
\end{eqnarray}
where $\eta=\ln{1\over x_B}$ and $z_{12}\equiv z_1-z_2$ etc. (we denote operators by ``hat'').
The first three terms in the BK equation correspond to the linear BFKL evolution \cite{bfkl} and describe the partons emission 
while the last term is responsible for the partons annihilation. For sufficiently low $x_B$ the partons emission 
balances the partons annihilation so the partons reach the state of saturation \cite{saturation} with
the characteristic transverse momentum $Q_s$ growing with energy $1/x_B$
(for a review, see \cite{satreviews}).

\section{Next-to-leading order photon impact factor}

In the Regge limit all transverse momenta
are of the same order of magnitude and consequently it is natural to introduce a factorization scale
in rapidity: one introduces a rapidity divide $\eta$ which separate ``fast'' field from ``slow'' fields. Thus,
the amplitude of the process is given by
a convolution of contributions coming from fields with rapidity $\eta<Y$ (``fast'' field) and contributions coming from fields
with rapidity $\eta>Y$ (``slow'' fields). As in the case of the usual Operator Product Expansion (OPE), 
the integration over the field with rapidity $\eta<Y$ 
gives us the coefficients function while the integrations over the field with rapidity $\eta>Y$ are the matrix elements of the operators.
Thus, the OPE at high-energy (Regge limit) for the T-product of two electromagnetic currents
is obtained in terms of Wilson lines
\begin{eqnarray}
&&\hspace{-8mm}
T\{\hat{j}_\mu(x)\hat{j}_\nu(y)\}=\int\! d^2z_1d^2z_2~I^{\rm LO}_{\mu\nu}(z_1,z_2)
[{\rm Tr}\{\hat{U}^\eta_{z_1}\hat{U}^{\dagger\eta}_{z_2}\}]^{\rm conf}
\nonumber\\
&&\hspace{-8mm}
+\int\! d^2z_1d^2z_2d^2z_3~I^{\rm NLO}_{\mu\nu}(z_1,z_2,z_3)
[ {\rm tr}\{\hat{U}^\eta_{z_1}\hat{U}^{\dagger\eta}_{z_3}\}{\rm tr}\{\hat{U}^\eta_{z_3}\hat{U}^{\dagger\eta}_{z_2}\}
-N_c{\rm tr}\{\hat{U}^\eta_{z_1}\hat{U}^{\dagger\eta}_{z_2}\}]
\label{high-energy-ex}
\end{eqnarray}
where
\begin{eqnarray}
&&\hspace{-15mm}
[{\rm Tr}\{\hat{U}^\eta_{z_1}\hat{U}^{\dagger\eta}_{z_2}\}\big]^{\rm conf}={\rm Tr}\{\hat{U}^\eta_{z_1}\hat{U}^{\dagger\eta}_{z_2}\}\nonumber\\
&&\hspace{-15mm}
+{\alpha_s\over 2\pi^2}\!\int\! d^2 z_3~{z_{12}^2\over z_{13}^2z_{23}^2}
[ {\rm Tr}\{T^n\hat{U}^\eta_{z_1}\hat{U}^{\dagger\eta}_{z_3}T^n\hat{U}^\eta_{z_3}\hat{U}^{\dagger\eta}_{z_2}\}
-N_c {\rm Tr}\{\hat{U}^\eta_{z_1}\hat{U}^{\dagger\eta}_{z_2}\}]
\ln {az_{12}^2\over z_{13}^2z_{23}^2}
\label{confodipole}
\end{eqnarray}
\begin{figure}
\centering
\includegraphics[width=0.8\textwidth]{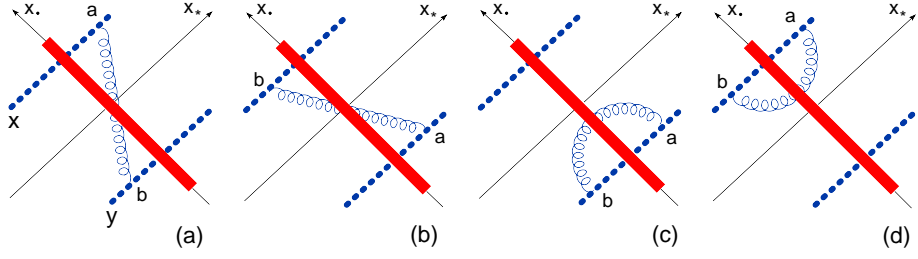}
\caption{Leading-order diagrams for the small-$x$ evolution of color dipole\label{bkevol}. Wilson lines are denoted by dotted lines.}
\end{figure}
is the \textit{composite dipole} with the conformal longitudinal cutoff in the next-to-leading order. The appearance of 
the \textit{composite operators} is due to the loss of conformal invariance of the Wilson line operator in the NLO.
Indeed, the light-like Wilson lines $U(x_\perp)$ are formally M\"obius invariant and consequently 
the leading-order BK equation is also conformal invariant. 
At NLO the Wilson line operator are divergent and its regularization introduces a dependence on the rapidity and conformal 
symmetry is lost. In order to restore the conformal invariance we redefine the operator ${\rm Tr}\{\hat{U}^\eta_{z_1}\hat{U}^{\dagger\eta}_{z_2}\}$ 
by adding suitable \textit{conterterms}. The procedure of finding the dipole with conformally regularized rapidity divergence is analogous
to the construction of the composite renormalized local operator by adding the appropriate counterterms order by order in perturbation theory.
In equation (\ref{high-energy-ex}) the coefficient $I^{\rm LO}$ is the leading-order (LO) impact factor which has been known for long time, while 
$I^{\rm NLO}$ is the NLO impact factor \cite{NLOIF} given by
\begin{eqnarray}
\hspace{-0.5cm}&&I^{NLO}_{\mu\nu}(x,y)=
- {\alpha_s N_c^2\over 8\pi^7 x_*^2y_*^2}\!\int\! d^2z_1d^2z_2 ~\calu^{\rm conf}(z_1,z_2)\Bigg\{\Bigg[
{1\over \calz_1^2\calz_2^2}\partial^x_\mu\partial^y_\nu\ln{\Delta^2\over x_\ast y_\ast}
\label{NLOIF}
\\
\hspace{-0.5cm}&&
+~2{\big(\partial_\mu^x\calz_1\big)\big(\calz_2\partial_\nu^y\big)\over \calz_1^3\calz_1^3}
\big[\ln{1\over R}+{1\over 2R}-2\big]
+~{2\big(\partial^x_\mu\calz_1\big)\big(\partial^y_\nu\calz_1\big)\over\calz_1^4\calz_2^2}\big[\ln{1\over R}-{1\over 2R}\big]
\nonumber\\
\hspace{-0.5cm}&&
-~{1\over 2}\Big[
{\partial^x_\mu\calz_1\over\calz_1^3 \calz_2^2}\partial^y_\nu\ln{\Delta^2\over x_\ast y_\ast}
+{\partial^y_\nu\calz_1\over \calz_1^3\calz_2^2}\partial^x_\mu\ln{\Delta^2\over x_\ast y_\ast}
\Big]\big(1-{1\over R}\big)
-{1\over 2\calz_2^2}\Big[\big(\partial^x_\mu{1\over \calz_1^2}\big)\partial^y_\nu R+\big(\partial^y_\nu{1\over \calz_1^2}\big)\partial^x_\mu R\Big]
{\ln R\over 1-R}   
\nonumber\\
\hspace{-0.5cm}&&
-~\big(\partial^x_\mu\partial^y_\nu{\Delta^2\over x_\ast y_\ast}\big){R^3\over z_{12}^4}
\big[{1\over R}+{3\over 2R^2}-2\big]\big({x_\ast y_\ast\over\Delta^2}\big)^3
+{1\over R}\Big[ {\partial^x_\mu\calz_1 \over \calz_1^3\calz^2_2}\big(\partial^y_\nu\ln{\Delta^2\over x_\ast y_\ast}\big)
+{\partial^y_\nu\calz_1\over\calz_1^3\calz^2_2}\big(\partial^x_\mu\ln{\Delta^2\over x_\ast y_\ast}\big)\Big]             
\nonumber\\
\hspace{-0.5cm}&&
+ ~4 {\big(\partial^x_\mu\calz_1\big)\big(\partial^y_\nu\calz_2\big)\over \calz_1^3\calz_2^3}\Big[
4{\rm Li}_2(1-R)-{2\pi^2\over 3}
+2(\ln R-1)\big(\ln R-{1\over R}\big)\Big]
\nonumber\\
\hspace{-0.5cm}&&
+~2{\big(\partial^x_\mu\calz_1\big)\big(\partial^y_\nu\calz_2\big)\over \calz_1^3\calz_2^3}\Big[{\ln R\over R(1-R)}-{1\over R}+2\ln R -4\Big]
+2{\big(\partial^x_\mu\calz_1\big)\big(\partial^y_\nu\calz_1\big)\over \calz_1^4\calz_2^2}\Big[{\ln R\over R(1-R)}-{1\over R}\big]
\nonumber\\
\hspace{-0.5cm}&&
-\Big({\partial^x_\mu\calz_1\over\calz_1^3\calz_2^2}\partial^y_\nu\ln{\Delta^2\over x_*y_*}
+{\partial^y_\nu\calz_2\over\calz_2^3\calz_1^2}\partial^x_\mu\ln{\Delta^2\over x_*y_*}\Big) 
\Big[{\ln R\over R(1-R)}-2\Big]+~(z_1\leftrightarrow z_2)\Bigg]
\nonumber\\
\hspace{-0.5cm}&&
-~2{z_{12\perp}^2\over \calz_1^3\calz_2^3}\Big[4{\rm Li}_2(1-R)-{2\pi^2\over 3}
+2(\ln {1\over  R}+{1\over R}+{1\over 2R^2}-3)\ln{1\over R}-\big(6+{1\over R}\big)\ln R+{3\over R}
-4\Big]\partial^x_\mu\partial^y_\nu{\Delta^2\over x_\ast y_\ast}\Bigg\}\nonumber
\end{eqnarray}
where 
\begin{eqnarray}
&&\Delta\equiv (x-y),~~~~~~~~x_*= x^+\sqrt{s/2},~~~~~~~~y_*=x^+\sqrt{s/2},~~~~~~~~
R\equiv -{\Delta^2z^2_{12\perp}\over x_*y_*\calz_1\calz_2}\nonumber\\
&&\calz_1=-{(x-z_1)^2\over x_*} + {(y-z_1)^2\over y_*}, ~~~~~\calz_2=-{(x-z_2)^2\over x_*} + {(y-z_2)^2\over y_*}
\end{eqnarray}
\begin{figure}
\begin{center}
\hspace{-0.5cm}
\includegraphics[width=29mm]{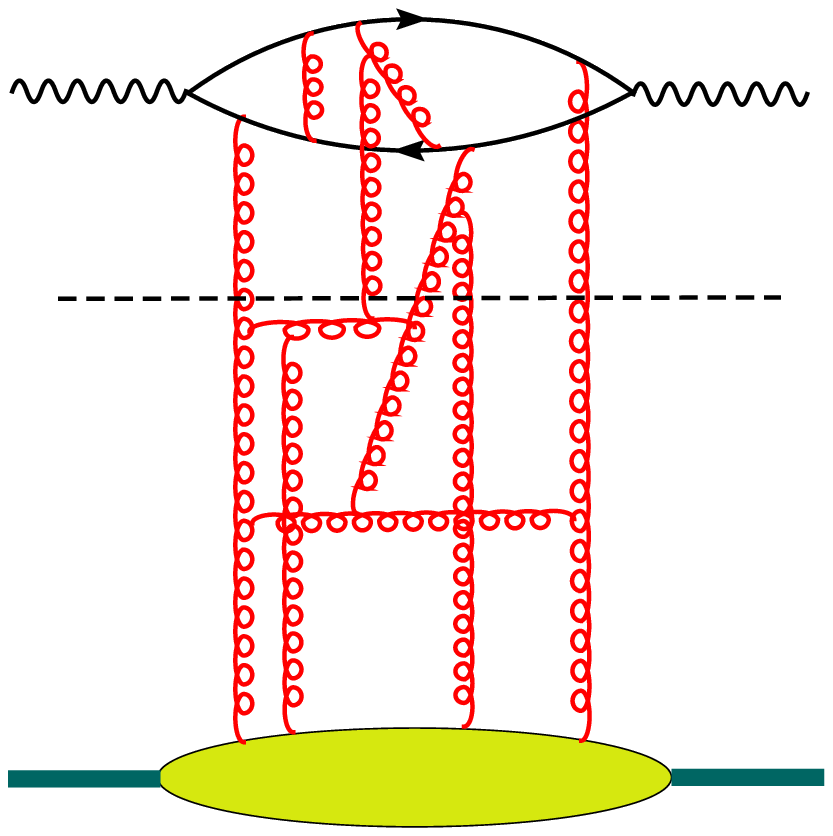}
\hspace{0.7cm}
\includegraphics[width=13mm]{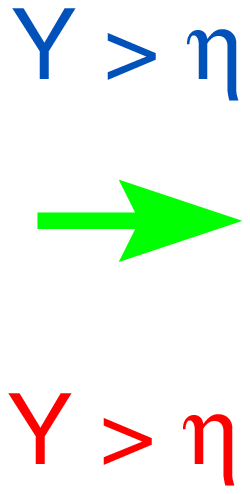}
\hspace{0.7cm}
\includegraphics[width=54mm]{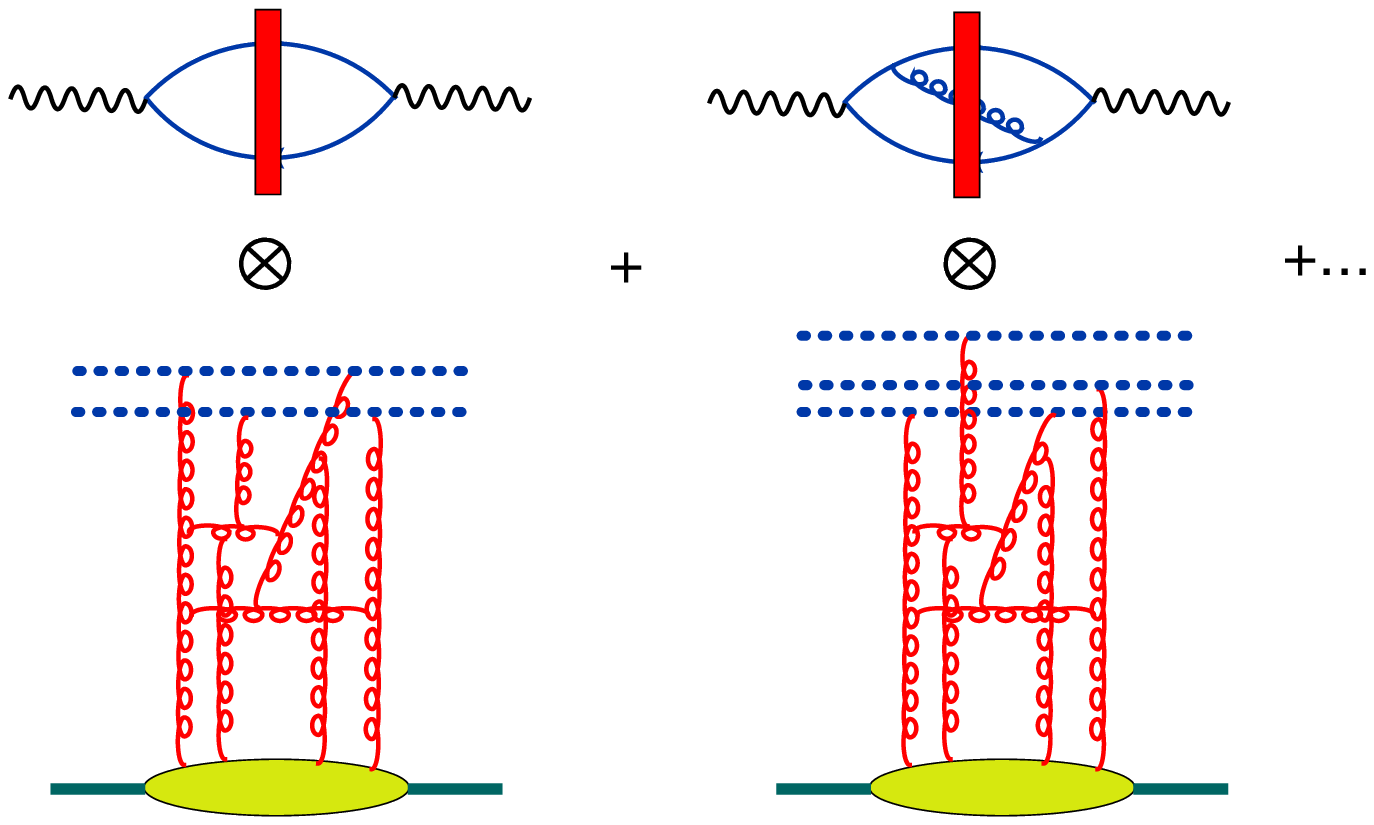}
\label{he-ex}
\end{center}
\caption{High energy expansion of the ${\rm T}$ product of two electromagnetic currents}
\end{figure}
Equation (\ref{NLOIF}) is the analytic expression for the full NLO impact factor which was not known before. (A combination of 
numerical and analytical results can be found in Ref. \cite{bart1}.) 
We plan to perform the Fourier transform in momentum space which will be useful for phenomenological studies.

In order to obtain the NLO evolution for the DIS amplitude in QCD one needs the NLO
evolution equation of color dipoles with respect to rapidity which was found in \cite{nlobk}, then solve the corresponding evolution equation,
and finally assemble the result for structure functions: take the initial conditions at low energy (rapidity), evolve color dipoles to higher
rapidity and multiply the result by the corresponding impact factor. The work is in progress.

In Ref. \cite{nlobkN4} the full program for the calculation of the 
NLO evolution amplitude is performed for the ${\cal N}=4$ SYM theory for two BPS-protected currents.

The author is grateful to the organizers of DIS 2010 and in particular to D. Colferai for financial support. 
This work is supported by the grant ANR-06-JCJC-0084.

\end{document}